\documentclass[a4paper,11pt]{article}
\usepackage{jheppub} 
\usepackage{lineno}
\DeclareMathAlphabet\bfcal{OMS}{cmsy}{b}{n}

\usepackage{dcolumn}
\usepackage[normalem]{ulem}
\usepackage{bm}
\usepackage{epsfig}
\usepackage{graphicx}
\usepackage{amsmath}
\usepackage{amssymb}
\usepackage{mathrsfs}
\usepackage{color}
\usepackage[usenames,dvipsnames]{xcolor}
\usepackage{hyperref}
\usepackage{feynmp-auto}
\usepackage{tikz}
\usepackage{braket}
\usepackage[caption=false]{subfig}
\usepackage{multirow}
\usepackage{array}
\usepackage{slashed}
\usepackage{rotating}
\usepackage{caption}
\usepackage{subcaption}
\usepackage[T1]{fontenc}

\arxivnumber{2508.00814} 

\newcommand{\rmd}{{\rm{d}}}

\newcommand\identity{1\kern-0.25em\text{l}}

\newcommand{\Otsosing}{\braket{\mathcal{O}^{\psi(2S)}(^3S_1^{[1]})}}
\newcommand{\Otsooct}{\braket{\mathcal{O}^{\psi(2S)}(^3S_1^{[8]})}}
\newcommand{\Oosz}{\braket{\mathcal{O}^{\psi(2S)}(^1S_0^{[8]})}}
\newcommand{\Otpz}{\braket{\mathcal{O}^{\psi(2S)}(^3P_0^{[8]})}}

\title{\boldmath $\psi(2S)$ production in jets using NRQCD}

\author[a]{Marston Copeland,}
\author[b]{Lin Dai,}
\author[a]{Yu Fu,}
\author[a]{Jyotirmoy Roy}

\affiliation[a]{Department of Physics, Duke University, Durham, NC 27708, USA}
\affiliation[b]{Guangxi Key Laboratory for Relativistic Astrophysics, School of Physical Science and Technology, Guangxi University, Nanning 530004, P. R. China}

\emailAdd{paul.copeland@duke.edu}
\emailAdd{dailin@gxu.edu.cn}
\emailAdd{yu.fu@duke.edu}
\emailAdd{jyotirmoy.roy@duke.edu}

\abstract{Based on recent data from LHCb, we study $\psi(2S)$ production in jets using non-relativistic QCD (NRQCD) in conjunction with the Fragmenting Jet Function (FJF) and Gluon Fragmentation Improved \textsc{Pythia} (GFIP) formalisms. Similar to previous studies of $J/\psi$ production in jets, our results show that these formalisms offer a much better description of data than the default \textsc{Pythia}+NRQCD prediction. We compare and contrast between the predictions from the FJF formalism and the GFIP approach. In addition, our results show that the distribution of $\psi(2S)$ in jets is an excellent discriminator to test different predictions for the $\psi(2S)$ LDMEs from various extractions. We find a large disparity between the predictions from three different collaborations showing that a more precise extraction of the $\psi(2S)$ LDMEs may be necessary.}

\begin{document}
\maketitle
\flushbottom

\section{Introduction}
Quarkonia, bound states of heavy quarks, are rich systems that provide insight into both the perturbative and non-perturbative dynamics of QCD. This is facilitated by the fact that quarkonium dynamics encompass a wide range of disparate scales -- the mass of the heavy quark/anti-quark, the relative momentum of the heavy quark--anti-quark pair, the binding energy of the pair, and the hadronization scale. This scale separation warrants the application of effective field theory methods. Non-relativistic QCD (NRQCD) is one such effective field theory that disentangles the dynamics at different scales using a systematic expansion in $v$, the relative velocity of the heavy quark--anti-quark pair (see ref. \cite{Hoang:2002ae, Brambilla:2010cs} for a review). NRQCD has been quite successful in describing quarkonium production for many processes, \cite{Fleming:1997fq, Yuan:2000cn, Beneke:1998re, Chu:2024fpo, Maxia:2024cjh, Bodwin:2010fi, Braaten:2002fi, Hagiwara:2003cw, Bodwin:2008nf, Braaten:1994xb, Braaten:1994vv, Copeland:2023wbu,Copeland:2023qed, Blask:2025jua, Echevarria:2024idp, Echevarria:2023dme, Yang:2024ejk}. In particular, it has allowed one to calculate quarkonium production cross-sections by factorizing them into short distance partonic cross-sections that describe the production of the heavy quark--anti-quark pair at the hard scale, $2m_Q$, and long distance matrix elements (LDME) for the pair to hadronize to the final state quarkonium. While the short distance cross-section is perturbatively calculable, the LDMEs are non-perturbative and must be extracted from experiments. Thus a major effort in understanding quarkonium production has been directed towards constraining the LDMEs. Most of the LDME extractions come from inclusive production of quarkonium \cite{Butenschoen:2011yh,Ma:2010yw,Brambilla:2022rjd,Brambilla:2024iqg} and thus a broader class of observables are wanted for better constraining of the LDMEs.

In \cite{Baumgart:2014upa, Bain:2016clc}, it was realized that studying $J/\psi$ production in jets offered a new avenue to constrain the LDMEs used for $J/\psi$ production. This was achieved by combining the Fragmenting Jet function (FJF) Formalism, developed in ref. \cite{Procura:2009vm} to study light hadron production in jets, with NRQCD to study $J/\psi$ mesons produced within a jet. This work culminated in a successful phenomenological description of in-jet $J/\psi$ production data from the LHCb collaboration \cite{Bain:2017wvk} and polarization studies of quarkonia \cite{Kang:2017yde, Dai:2017cjq}, showing that this approach has considerable merit.

The vast majority of NRQCD applications to charmonium production have focused on the lowest lying state, the $J/\psi$ meson. This is in large part due to its extremely clean decay signal to lepton pairs, which makes it appealing to measure experimentally. Many attempts have been made to constrain the long-distance matrix elements for the $J/\psi$ \cite{Butenschoen:2011yh,Butenschoen:2012qr,Chao:2012iv,Bodwin:2014gia}. The first excited state of the $J/\psi$, the $\psi(2S)$ meson, played a critical role in the development of NRQCD however, when the unexpected surplus of $\psi(2S)$ mesons at the Tevatron lead to the understanding that fragmentation and so-called ``color-octet'' production mechanisms can contribute huge amounts to the overall production cross sections in NRQCD \cite{Braaten:1994vv, Braaten:1994xb}. Over time though, studies of the $\psi(2S)$ meson have become less popular, and as a result, its long-distance matrix elements are poorly constrained. The $\psi(2S)$'s most common decay is to a $J/\psi$ and a $\pi^+\pi^-$ pair, so recently several studies have recognized the importance of $\psi(2S)$ decay contributions in constraining $J/\psi$ production physics \cite{Bodwin:2015iua}. Compared to the $J/\psi$, $\psi(2S)$ production offers a particularly clean testing ground for NRQCD fragmentation mechanisms, as its production from the feeddown of higher charmonium resonances is significantly smaller compared to $J/\psi$ \cite{Faccioli:2008ir,LHCb:2012geo,ATLAS:2014zpz}. Several authors have begun to study the $\psi(2S)$ LDMEs with the same vigor as was applied to the $J/\psi$, \cite{Bodwin:2015iua, Brambilla:2022ayc, Butenschoen:2022qka}, but further work is still necessary to constrain the LDMEs for the $\psi(2S)$.

In the present work, we will study $\psi(2S)$ production in jets based on the FJF framework and Gluon Fragmentation Improved \textsc{Pythia} (GFIP). This is motivated by the recent LHCb measurements of $\psi(2S)$ meson production in jets \cite{LHCb:2024ybz}. In their analysis, the $\psi(2S)$ is reconstructed via its decay to the $J/\psi(\rightarrow \mu^+ \mu^-) \pi^+ \pi^-$ final state in the forward region, using proton-proton collision data collected by the LHCb experiment at a center-of-mass energy of 13\,TeV in 2016. They measured the transverse momentum distribution of $\psi(2S)$ inside jets {constructed via the \texttt{anti-$k_T$} jet algorithm}, defining it as the ratio of the transverse momentum of the $\psi(2S)$ candidate to that of the full jet, differentially in both $p_T(\mathrm{jet})$ and $p_T(\psi(2S))$ bins. They also provided a comparison with theoretical predictions from a combined NRQCD and \textsc{Pythia} model, which showed that a naive combination of NRQCD with \textsc{Pythia} fails catastrophically. Similar issues were also faced in the case of $J/\psi$ \cite{LHCb:2017llq}. Monte Carlo (\textsc{Pythia}) was initially used to interpret the data for $J/\psi$ distribution in jets at LHCb and was not able to explain the data. In order to mend the insufficient part of \textsc{Pythia}, and motivated by FJF framework, GFIP was introduced in Ref.~\cite{Bain:2016clc, Bain:2017wvk} that still used \textsc{Pythia} to do the parton shower part, but then the hadronization part was manually implemented. Theoretically, the FJF and GFIP approaches should be equivalent, since renormalization group (RG) evolution in FJF is equivalent to parton shower in GFIP.

The structure of the paper is as follows. Sec. \ref{sec:formalism} as a whole reviews the various formalisms used in this work with Secs. \ref{sec:NRQCD}, \ref{sec:FJFformalism}, \ref{sec:GFIP} discussing NRQCD factorization, FJF and GFIP formalisms respectively. In Sec. \ref{sec:Results} we show the results of FJF and GFIP for different LDMEs and compare them to LHCb data. Finally, we summarize our work in Sec. \ref{sec:Conclusion}.

\section{Formalism Review}
\label{sec:formalism}

%
\subsection{NRQCD Factorization Formalism}
\label{sec:NRQCD}

One of the major tools that allows for NRQCD to have predictive power in quarkonium production processes is the NRQCD factorization conjecture. This states that for the collision process of particles $A$ and $B$, the production of a quarkonium particle, $H$ and other final states, $X$, can be factorized as follows, 
\begin{equation}
\label{eq: NRQCD fact}
    \rmd\sigma_{A+B\to H+X}\ = \sum_n \rmd\sigma_{A+B\to Q\bar{Q}(n)+X} \braket{\mathcal{O}^{H}(n)} \, .
\end{equation}
Here, $d\sigma_{A+B\to Q\bar{Q}(n)+X}$ is a partonic cross section that describes the production of a $Q\bar{Q}$ pair in the quantum number configuration denoted by $n= {}^{2S+1}L_J^{[c]}$ for definite spin $S$, orbital
angular momentum $L$, total angular momentum $J$, and color state $c$, where $c=1$ for color-singlet and $c=8$ for color-octet. These partonic cross sections are perturbatively calculable at the scale $2m_Q$ in a power series of $\alpha_s$. The long-distance matrix elements (LDMEs), $\braket{\mathcal{O}^{H}(n)}$, are vacuum production matrix elements of operators describing the transition of the heavy quarks in quantum number configuration $n$ into the final bound state quarkonium $H$. These LDMEs can be treated as nonperturbative parameters that can be extracted from experiment, and they also scale with definite powers of the relative velocity $v$ of the heavy quarks. Consequently, the NRQCD factorization framework systematically organizes quarkonium production as a double expansion in the strong coupling constant $\alpha_s$ and relative velocity $v$.

For the $\psi(2S)$, the relevant operators are essentially the same as those needed to describe $J/\psi$ production. The excited nature of the $\psi(2S)$ does not affect the process of hadronization. It only modifies the size of the relative velocity, which is the power-counting parameter of the theory. The overall $v$-scaling of each matrix element remains unchanged and will be the same as the $v$-scaling for the matrix elements of the ground state $J/\psi$. The dominant production operator is given by $\Otsosing$, which scales as $v^3$, and there are three subleading operators denoted by $\Oosz$, $\Otpz$, $\Otsooct$, which all scale as $v^7$. These matrix elements have been extracted in several different studies, see refs. \cite{Bodwin:2015iua, Butenschoen:2022qka, Brambilla:2022ayc}. The results from these works are summarized in Table \ref{tab: LDMEs}

\begin{table}[htbp]
\centering
\begin{tabular}{c|c|c|c|c}
\hline
& $\begin{aligned}\Otsosing \\ \times \, \text{GeV}^3\end{aligned}$ & $\begin{aligned}\Otsooct \\ \times 10^{-2} \, \text{GeV}^3\end{aligned}$ & $\begin{aligned}\Oosz \\ \times 10^{-2} \, \text{GeV}^3\end{aligned}$ & $\begin{aligned}\Otpz/m_c^2 \\ \times 10^{-2} \, \text{GeV}^3\end{aligned}$\\
\hline
Bodwin et al.~\cite{Bodwin:2015iua} & $0.76 $ & $-0.157\pm 0.28 $ & $3.14 \pm 0.79$ & $-0.114 \pm 0.121$ \\
\hline
B\&K~\cite{Butenschoen:2022qka} & $0.76 $ & $0.276\pm 0.01 $ & $ 0.835\pm 0.096$ & $0.865 \pm 0.055 $ \\
\hline
Brambilla et al. ~\cite{Brambilla:2022ayc} & $0.71 \pm 0.21 $ & $0.84\pm 0.25 $ & $ -0.37 \pm 1.92$ & $ 1.55 \pm 4.9 $ \\
\hline
\end{tabular}
\caption{$\psi(2S)$ NRQCD LDMEs.\label{tab: LDMEs}}
\end{table}
%


Notice in many of these studies, the color octet LDMEs have uncertainties on the order of 100-200$\%$. For the $^3S_1^{[8]}$ and $^3P_J^{[8]}$, this could be due to large cancellations that occur between the $^3S_1^{[8]}$ and $^3P_J^{[8]}$ partonic cross sections in $pp$ collisions. However, in general, this is indicative that more data is necessary to constrain these parameters. One of the purposes of this work is to investigate these matrix elements by comparison of $\psi(2S)$ production data in jets.

\subsection{NRQCD Fragmenting Jet Function Formalism}\label{sec:FJFformalism}
Theoretical studies of hadron production inside jets can be performed via analytical calculations based on factorization theorems. In particular, within the framework of Soft-Collinear Effective Theory (SCET), rigorous factorization theorems have been established that separate the dynamics at different energy scales, allowing for systematic calculations of jet substructure observables. These formalisms, called ``Fragmenting Jet Function Formalisms”, enable a perturbative description of the hard scattering and parton shower, matched to nonperturbative functions that describe hadronization inside jets. One of the core objects in this formalism is the fragmenting jet function (FJF).

The FJF formalism can be applied to many different production processes with a quarkonium inside a jet. In the case of our present interest, we consider proton-proton collisions at a center-of-mass energy of 13 TeV. For the production cross section for a jet with energy $E$, cone size $R$, and a $\psi(2S)$ meson with energy fraction $z$, the factorization theorem is schematically of the form
\begin{equation}
    \frac{\rmd \sigma}{\rmd E \rmd z} =\sum_{a,b}\sum_{i,j}  f_{a/p}\otimes f_{b/p}\otimes H_{ab\to ij} \otimes J_j\otimes S\times {\cal G}^{\psi(2S)}_i(E,R,z,\mu) 
\label{eq: FJF factorization}
\end{equation}
where $f_{a/p}$ and $f_{b/p}$ are parton distribution functions, $H_{ab\to ij}$ is the hard coefficient, $J_j$ represents the jet function for a jet without $\psi(2S)$ constituents initiated by a final state parton $j$, ${\cal G}^{\psi(2S)}_i(E,R,z,\mu)$  is the FJF for the jet containing the $\psi(2S)$ fragmenting from parton $i$, and $S$ is the soft function. All of the $z$ dependence is contained in ${\cal G}^{\psi(2S)}_i(E,R,z,\mu)$. In general, for any hadron, $H$, the FJF has a factorized form as
\begin{equation}
    {\cal G}^H_i(E,R,z,\mu) =  \sum_{j} \int_z^1 \frac{\rmd z'}{z'}{\cal J}_{ij}(E,R,z',\mu)  D_{j\to H}(z/z',\mu)  +{\cal O}\left(\Lambda_{\rm QCD}^2/E^2R^2\right),
    \label{eq:FJF}
\end{equation}
where $z$ is the fraction of energy carried by the hadron in the jet, $E$ is the jet energy, and $R$ is the jet radius. ${\cal J}_{ij}(E,R,z',\mu)$ is the matching coefficient, which is calculable using perturbative QCD\footnote{{In this paper, we use the matching coefficients calculated at NLO in $\alpha_s$ for anti-$k_T$ jets, which have already been calculated in \cite{Waalewijn:2012sv}.}}, and the $D_{j\to H}$ is the fragmentation function. In the study of quarkonium production inside jets, Ref.\cite{Baumgart:2014upa} showed that the FJF for quarkonium is still of the convolution form in eq. (\ref{eq:FJF}) and they calculated the FJF in terms of NRQCD fragmentation functions. Since the matching coefficient ${\cal J}_{ij}(E,R,z',\mu)$ does not depend on the particular hadron species observed in the final state, for quarkonium, these coefficients are still the same as those for light hadrons, which have already been calculated in ref. \cite{Procura:2011aq}. What makes the FJF for light hadrons different from that for quarkonium is the fragmentation function $D_{j\to H}$. 

For heavy charmonium states, like the $\psi(2S)$, the fragmentation functions can be computed by matching onto the LDMEs at the scale $2m_c$,
\begin{equation}
    D_{i \to \psi(2S)} (z; \mu = 2m_c) = \sum_n d_{i\to c\bar{c}(n)}(z; \mu = 2m_c) \braket{\mathcal{O}^{\psi(2S)}(n)}
\end{equation}
where $d_{i\to c\bar{c}(n)}(z)$ is the perturbatively calculable partonic cross section for a parton fragmenting to a $c\bar{c}$ pair in the configuration $n$ and $\braket{\mathcal{O}^{\psi(2S)}(n)}$ are the LDMEs. The leading quantum number configurations for the $\psi(2S)$ and the corresponding LDMEs are discussed in the above section. 

The matching of the $\psi(2S)$ fragmentation functions should be almost identical to the procedure for matching the $J/\psi$'s fragmentation functions onto the $J/\psi$ LDMEs. That is, the perturbative matching coefficients, $d_{i\to c\bar{c}(n)}(z)$, for both charmonium states are identical. The matching of the $J/\psi$ fragmentation function onto each NRQCD mechanism can be found in ref. \cite{Baumgart:2014upa}, along with the full FJFs for the $J/\psi$ meson. Our results for the $\psi(2S)$ can be determined by plugging in the $\psi(2S)$ LDMEs to the expressions given in Appendix A of ref. \cite{Baumgart:2014upa},
\begin{equation}
    d_{i\to c\bar{c}(n)}(z) \braket{\mathcal{O}^{\psi(2S)}(n)} = D_{i\to \psi}^n(z) \bigg|_{\braket{\mathcal{O}^{\psi}(n)} = \braket{\mathcal{O}^{\psi(2S)}(n)}}.
\end{equation}

Since the NRQCD matching is computed at the scale $2m_c$, the resulting fragmentation function needs to be evolved to the factorization scale in the cross section. When using the FJF formalism to study quarkonium production in jets, we set the factorization scale in the jet function matching coefficient $ J_{gg}(E, R, z, \mu) $ to be $\mu_J = 2E\tan (R/2)$, which minimizes logarithms of $2E\tan (R/2)$. Therefore, for our purposes, the fragmentation function should be evolved from the initial scale $2m_c$ up to $\mu_J$. In the FJF formalism, we evolve the fragmentation function analytically using DGLAP evolution equations,
\begin{equation}
    \mu\frac{\partial}{\partial \mu}D_i^H(z,\mu)=\frac{\alpha_s(\mu)}{\pi} \sum_{j}\int_z^1\frac{\rmd z'}{z'} P_{i\to j}(\frac{z}{z'},\mu)D_j^H(z',\mu).
    \label{eq:FFDGLAP}
\end{equation}
We solve these differential equations in eq. (\ref{eq:FFDGLAP}) in Mellin space, where the convolution becomes a product, and then return to momentum space by taking an inverse Mellin transform numerically. 

At the cross-section level, along with other jet functions and soft functions, the fragmenting jet function will be convolved with the hard function, as in eq. (\ref{eq: FJF factorization}). All of these functions are evaluated at their characteristic scale, and then using renormalization group techniques, they are evolved to a common hard scale $\mu$ so that large logarithms in each component will be resummed to all orders in perturbation theory. Details of this formalism for the calculation of jet cross sections accurate to next-to-leading-log can be found in the seminal paper \cite{Ellis:2010rwa}, and an analytical calculation for jets containing a B meson or $J/\psi$ was performed at next-to-leading-log-prime(NLL') precision in Ref.\cite{Bain:2016clc}, which goes beyond standard NLL accuracy by incorporating the full set of $\mathcal{O}(\alpha_s)$ fixed-order corrections to all relevant functions. 

\subsection{Gluon fragmentation improved \textsc{Pythia}}
\label{sec:GFIP}
\begin{figure}
    \centering
    \includegraphics[width=\linewidth]{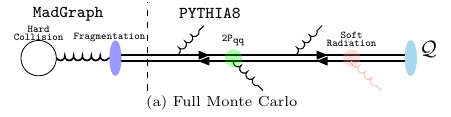}\\~\\

    \includegraphics[width=\linewidth]{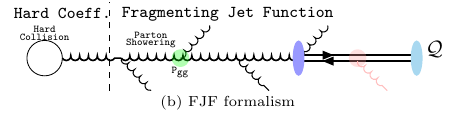}\\~\\

    \includegraphics[width=\linewidth]{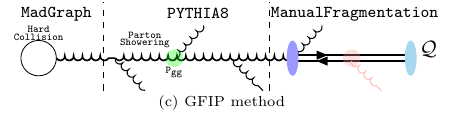}
    \caption{Comparison of different theoretical treatments for quarkonium production inside jets.}
    \label{fig:workflow}
\end{figure}
Theoretical studies of hadron production inside jets can also be carried out using Monte Carlo methods, with a common approach involving the use of the \textsc{Pythia} event generator\cite{Bierlich:2022pfr}. In the case of quarkonium production within jets, it has been shown that the default \textsc{Pythia} setup fails to describe the corresponding data for both $J/\psi$~\cite{LHCb:2017llq} and $\psi(2S)$~\cite{LHCb:2024ybz}. This discrepancy is not unexpected, as the default \textsc{Pythia} modeling of quarkonium production does not capture the full complexity of quarkonium dynamics. 

When using \textsc{Pythia} to model the quarkonium production,  the $Q\bar{Q}$ pair is generated in either a color-singlet or color-octet configuration during the hard scattering process. If the pair is in a color-singlet state, it is assumed to hadronize directly into the quarkonium without emitting any QCD radiation, effectively behaving as a point-like color-neutral object. In contrast, for a color-octet configuration, the pair is treated as a single colored particle that undergoes parton showering governed by a splitting function $2P_{qq}(z)$, which strongly favors large momentum fractions $z \approx 1$. This full process is demonstrated in Fig.\ref{fig:workflow}(a). Consequently, the pair retains most of its momentum throughout the shower and ultimately transitions into a physical quarkonium by radiating a soft gluon to neutralize its color. 

The physics picture described above is different from the NLL' calculation within the FJF treatment, which is equivalent to producing a hard gluon in the short-distance process with virtuality at the jet energy scale, allowing that gluon to shower until a
gluon with virtuality $2m_Q$ hadronizes into the quarkonium. The properties of the latter can also be implemented in the Monte Carlo simulation by leveraging \textsc{Pythia}. The corresponding implementation was referred to as gluon fragmentation improved \textsc{Pythia}(GFIP), which has been sketched in Fig.\ref{fig:workflow}(c). Specifically, one can utilize \textsc{Pythia} to simulate events where a gluon is generated in the hard scattering, disable the default hadronization module, and allow the parton shower to evolve down to a scale near $2m_Q$. In \textsc{Pythia}, the lower cutoff scale of the parton shower is controlled by the parameter \texttt{TimeShower:pTmin}, which sets the minimum transverse momentum—and thus the minimal virtuality—of particles participating in the shower. The default value of this parameter is 0.4~GeV. In our setup, we modify it to 1.6~GeV, corresponding to a virtuality scale of approximately $2m_Q$, in order to align with the physical scale at which quarkonium hadronization is expected to occur. The resulting gluon energy fraction distribution can then be extracted and manually convolved with a perturbative NRQCD fragmentation function, computed at leading order in $\alpha_s(2m_Q)$. 
This procedure has been shown to yield good agreement with NLL' analytic calculations for the $z$-distribution of $J/\psi$ within jets \cite{Bain:2017wvk}.


\section{Results}
\label{sec:Results}
The measurement of $\psi(2S)$ production within fully reconstructed jets at the LHCb experiment at a center-of-mass energy of 13~TeV has been reported in~\cite{LHCb:2024ybz}. In this experimental study, the jet fragmentation function, defined as the distribution of the fraction of the transverse momentum, $z=p_T(\psi(2S))/p_T(\text{jet})$,
carried by the quarkonium-tag $\psi(2S)$ within a jet, was measured differentially in both $p_T(\text{jet})$ and $p_T(\psi(2S))$ bins. It was found that the prompt $\psi(2S)$ distributions differ significantly from fixed-order NRQCD predictions followed by a QCD parton shower using \textsc{Pythia} 8.309. To address this tension between experimental data and theoretical expectations, we calculate the jet fragmentation function of $\psi(2S)$ under the kinematic conditions of the LHCb measurement. We perform the calculation with two methods described in sec. \ref{sec:FJFformalism} and \ref{sec:GFIP}, namely, FJF  formalism and GFIP.

For the GFIP method, we first generate parton-level events corresponding to the hard production of charm quarks and gluons in proton-proton collisions at a center-of-mass energy of $\sqrt{s} = 13$~TeV using \textsc{MadGraph}\cite{Alwall:2014hca,Frederix:2018nkq}. The resulting parton-level events are passed to \textsc{Pythia} for parton showering, which is evolved down to a cutoff scale of approximately $2m_c$, as discussed earlier. After the parton shower, jets are reconstructed from the final-state particles using \texttt{anti-$k_T$} jet algorithm\cite{Cacciari:2008gp} implemented in $\textsc{FastJet}$ package\cite{Cacciari:2011ma}. To match the LHCb experimental setup, we impose jet requirements consistent with those used in Ref.~\cite{LHCb:2024ybz}, namely pseudorapidity $2.5 < \eta < 4.0$, jet radius $R = 0.5$, ${p_T}(\text{jet}) >5\text{GeV/c}$. Only charm quarks and gluons that are clustered into jets satisfying the aforementioned LHCb kinematic criteria are retained for further analysis. As has already been shown in Fig.(1) of Ref.\cite{Bain:2017wvk}, the resulting energy fraction $z$ distribution of charm quarks is sharply peaked near $z = 1$, while the gluon distribution is significantly softer, with a peak at lower $z$. 

The $p_T$ and $\eta$ distributions of charm quarks and gluons obtained from the simulation are subsequently convolved with the leading-order NRQCD fragmentation functions to compute the corresponding $p_T$ and $\eta$ distributions for $\psi(2S)$ production.  In the case of gluon fragmentation, we include the full set of leading NRQCD channels relevant for $\psi(2S)$ production: the color-singlet ${}^3S_1^{[1]}$ as well as the color-octet ${}^3S_1^{[8]}$, ${}^1S_0^{[8]}$, and ${}^3P_J^{[8]}$ contributions. Although color-octet LDMEs are suppressed by powers of the heavy quark relative velocity $v$, this suppression is offset by enhancements in the corresponding short-distance coefficients. For gluon fragmentation, the $^3S_1^{[8]}$ mechanism is leading in $\alpha_s$, the ${}^1S_0^{[8]}$ and ${}^3P_J^{[8]}$ channels appear at $\mathcal{O}(\alpha_s^2)$, and the ${}^3S_1^{[1]}$ contribution enters at $\mathcal{O}(\alpha_s^3)$. For charm quark fragmentation, both the color-singlet and color-octet channels arise at the same order in $\alpha_s$; however, given the relative size of the LDMEs, we retain only the color-singlet ${}^3S_1^{[1]}$ contribution in this case. The yielding $\psi(2S)$ mesons subsequently decay into $\pi^+ \pi^-$ and $J/\psi$, the latter finally decays into $\mu^+\mu^-$. We assume these decays are isotropic in the rest frame of the decaying systems. We supplement a detailed description of the four-body decay of $\psi(2S)$ in app.~\ref{App:A}. For both muon and pion candidates, their pseudorapidity must lie within the range $2.0 < \eta < 4.5$, and their transverse momenta must satisfy $p_T > 0.5~\text{GeV}/c$. In addition, muon candidates are required to have momentum $p(\mu) > 6~\text{GeV}/c$, while pion candidates must satisfy $p(\pi) > 3~\text{GeV}/c$. Finally, we obtain a normalized $z(\psi(2S))$ distribution for each production mechanism from parton $i$ to quantum state $n$. To ensure that the total number of $\psi(2S)$ coming from each mechanism is in the proper ratio, each mechanism is multiplied by the weight
\begin{equation}
    w(i,n)=\frac{\rmd \sigma(pp\to i+X)\int_0^1\rmd z D^n_{i\to\psi(2S)}(z)}{\rmd \sigma(pp\to c+X)\int_0^1\rmd z D^{{}^3S_1^{[1]}}_{c\to\psi(2S)}(z)}
\end{equation}

For calculation using FJF formalism, we obtained the FJFs for $\psi(2S)$ as the procedure explained in sec. \ref{sec:FJFformalism} for anti-$k_T$ jet with cone size $R=0.5$. The energy distributions of partons from the hadron collisions are initially obtained from \textsc{MadGraph}. As explained in sec. \ref{sec:FJFformalism}, we use DGLAP to evolve the NRQCD fragmentation functions corresponding to different fragmentation mechanisms from scale $\mu=2m_c$ to $\mu=p_T(\text{jet})$. We then convolve these results with the \textsc{MadGraph} energy distributions and weight them with the gluon and charm distributions. Finally, these results are convolved with the matching coefficients $\mathcal{J}_{ij}$ to obtain the $z(\psi(2S))$ distributions. Similar to GFIP the fragmentation functions of interest are the gluon color singlet ${}^3S_1^{[1]}$ and the color-octet ${}^3S_1^{[8]}$, ${}^1S_0^{[8]}$, and ${}^3P_J^{[8]}$ and the charm color singlet channel ${}^3S_1^{[1]}$ which mix with the color-singlet gluons in the DGLAP evolution. One additional thing to take into account in this calculation is the amount of $\psi(2S)$ that survives the muon and the pion cuts that we implement to reproduce the experimental measurement. Since these cuts were already performed in the GFIP analysis, we determine their effects on the final distributions by taking a ratio of the final GFIP results to the results before the kinematic cuts. We reproduce the effects of these kinematic cuts by fitting an analytic function to the output of the ratio, which we then use to normalize the FJF results. 

\begin{figure}[htp]
    \centering
    \includegraphics[width=\linewidth]{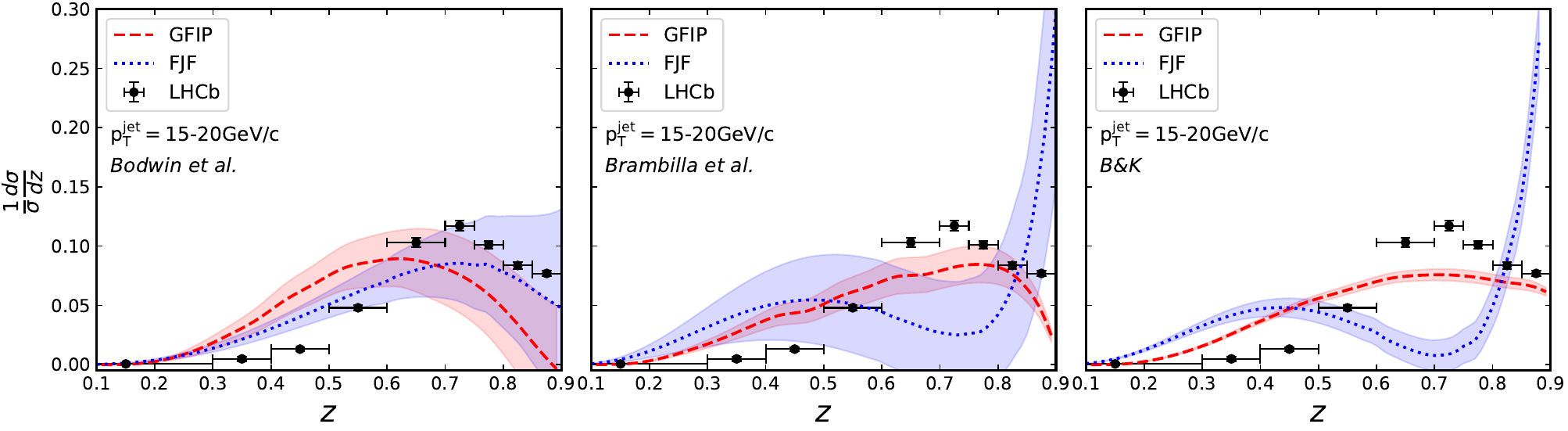}
    \includegraphics[width=\linewidth]{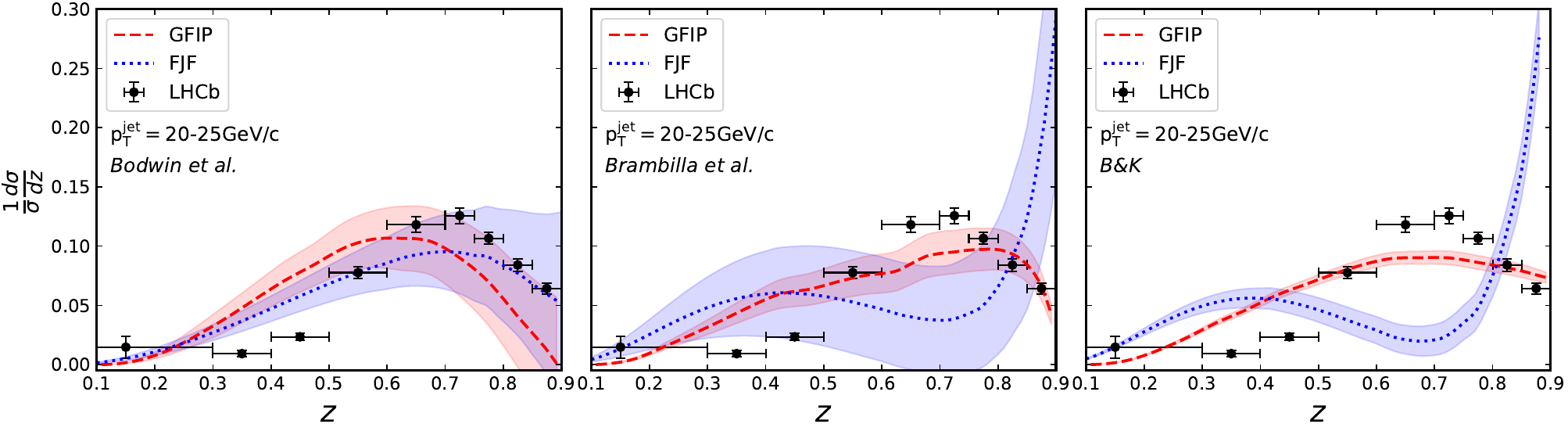}
    \includegraphics[width=\linewidth]{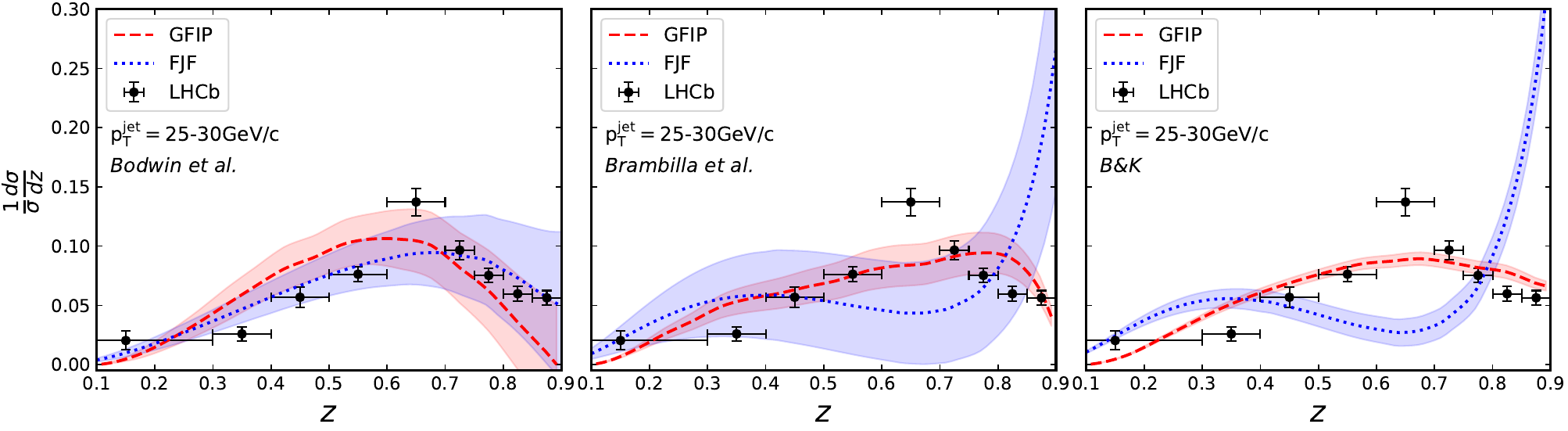}
    \includegraphics[width=\linewidth]{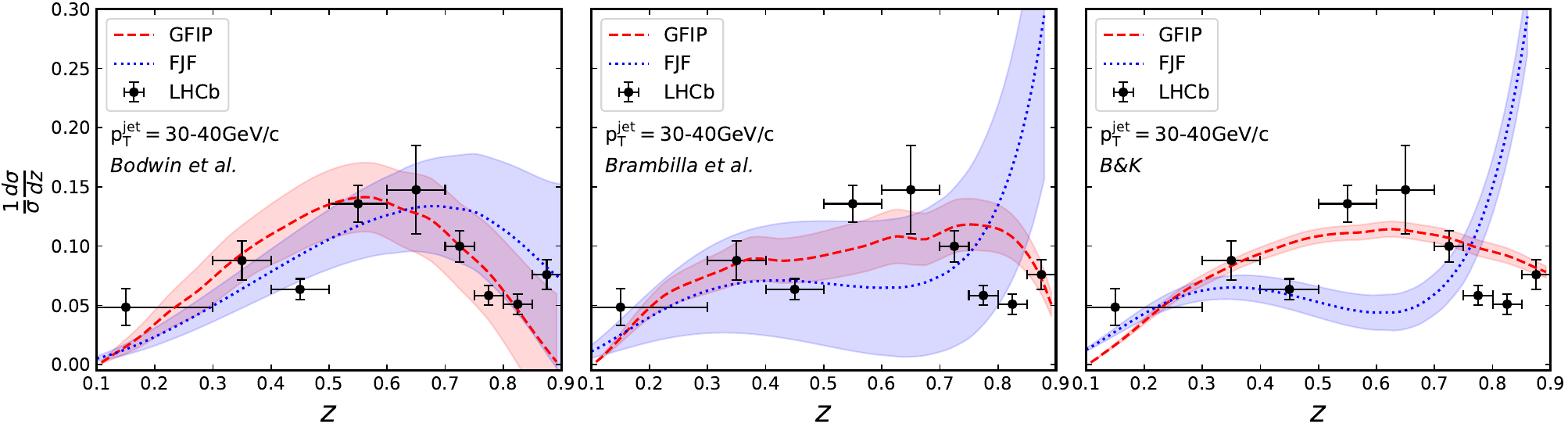}
    \includegraphics[width=\linewidth]{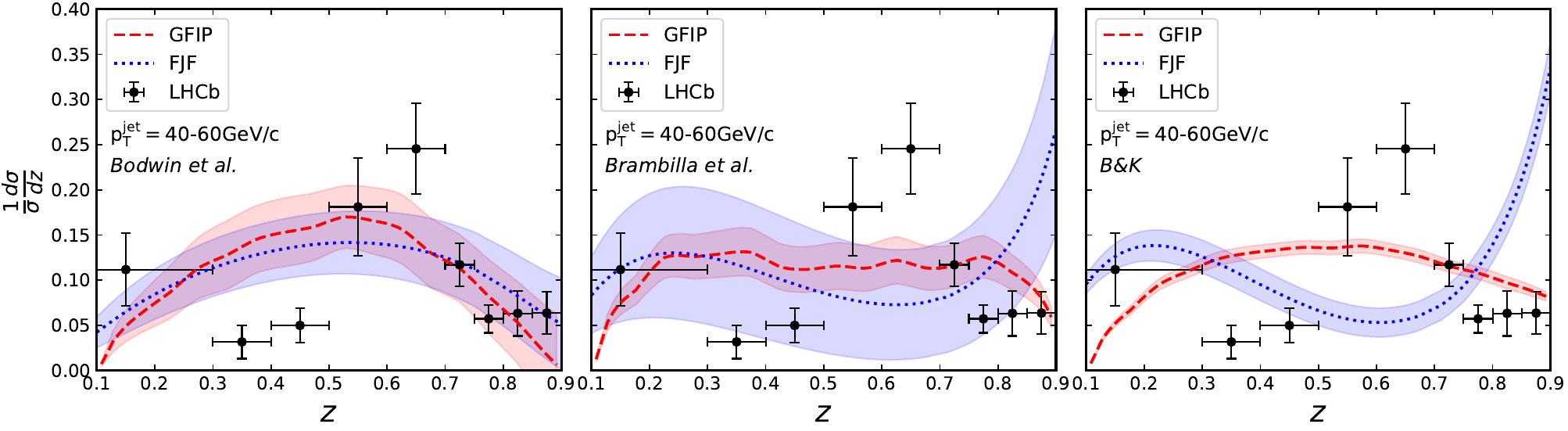}
    
    \caption{$z$-distributions of $\psi(2S)$ inside jets. Black dots are measured at the LHCb experiment with collision energy $\sqrt{s}=13$ TeV. The red and blue curves represent the results from the GFIP and FJF methods, respectively. Each row presents the results at a certain $p_T^{\text{jet}}$-bin with three different groups of LDMEs.}
    \label{fig: numerical results}
\end{figure}

In Fig. \ref{fig: numerical results}, we present our results for the $\psi(2S)$ distributions inside jets as a function of $z$, obtained using the two methods described above. The red curves represent the results from the GFIP approach, while the blue curves correspond to the calculations using the FJF method. These results are calculated using the three LDME sets listed in Table~\ref{tab: LDMEs}, including the extraction by Bodwin et al. \cite {Bodwin:2015iua}, Butenschoen and Kniel (B\&K) \cite{Butenschoen:2022qka}, and Brambilla et al. \cite{Brambilla:2022ayc}. The first to third columns correspond to the three different LDME sets, respectively, while each row shows results for a different $p_T^{\text{jet}}  $ bin. The shaded bands around the curves represent the uncertainty arising from the corresponding LDME errors. Other sources of uncertainty, such as scale variations, are not included. We compare with the corresponding LHCb experimental data at different jet ${p_T^{\text{jet}}}$ bins. To avoid regions where perturbative calculations may receive sizable corrections, particularly near the endpoints 
$z\to1$ and $z\to0$, we restrict our comparison with LHCb data to the intermediate range  $0.1<z<0.9$, and the resulting results are normalized to the total experimental yield in this window.

Overall, compared to the default \textsc{Pythia} simulations \cite{LHCb:2024ybz}, our calculations provide a significantly improved theoretical description of the LHCb measurements of $\psi(2S)$ distributions inside jets. This further supports the fragmenting jet formalism for quarkonium production, which has already demonstrated success in describing experimental results for $J/\psi$ \cite{LHCb:2017llq,Zhang:2024owr}. { In general, the predictions from each LDME set in the GFIP method agree quite well with the experimental measurements from the LHCb \cite{LHCb:2024ybz}.} The central values of the GFIP results generally lie close to the experimental data, except in the highest $p_T$(jet) bin, where the experimental points exhibit considerable scatter and are not described as well. { The narrative for the FJF predictions is slightly more complicated. The results using the Bodwin et al. matrix element set agree quite well with both the data and the analogous predictions from the GFIP formalism. However, for both the Brambilla et al. and B\&K sets, we observe significant discrepancies from the data and the GFIP predictions as $z\to 1$. This discrepancy becomes more dramatic as the $p_T$ bin increases. } Indeed, FJF and GFIP are not fully equivalent. In FJF, the NRQCD fragmentation function is evolved from $\mu = 2m_c$ to the jet scale $\mu_J$ and then convoluted with the matching coefficient (computed here at NLO), in which the term $\alpha_s(p_T^{\text{jet}}) \log(1 - z)$ diverges as $z\to1$. In GFIP, partons are showered in {\sc Pythia} from $\mu_J$ down to $\mu = 2m_c$ before being convoluted with the NRQCD fragmentation function. The two approaches are only approximately equivalent if the matching coefficient is taken at LO, where the FJF reduces to the standard fragmentation function. Even then, at LO the two approaches differ in practice near the endpoint region because the evolution is fundamentally different. PYTHIA evolves via a parton shower and orders the shower in relative transverse momentum ($k_T$) of the partons. Since $k_T \sim z(1-z) E_J \theta$, evolving down to $k_T \sim 2m_c$ does not effectively probe the $z \to 1$ region. By contrast, the DGLAP evolution in the FJF framework does access $z \to 1$, and is more akin to an angular ordering in the parton shower, rather than a $k_T$ ordering. As a result, the FJF prediction exhibits a significant enhancement of the cross section near the endpoint relative to GFIP, which we have explicitly confirmed already at LO, while {\sc Pythia} experiences an artificial suppression in this region.

Interestingly, only the Brambilla et al. and B$\&$K LDME sets suffer from these endpoint divergences, while the FJF results using the Bodwin et al. set appear relatively unchanged from the GFIP prediction. The reason this occurs for the Bodwin set and not the others is that, for the Bodwin et al. set, the contributions from the $^3P_J^{[8]}$ and $^3S_1^{[8]}$ channels (both of which diverge as $z\to1$) approximately cancel. Large cancellations between the $^3P_J^{[8]}$ and $^3S_1^{[8]}$ production channels have been observed in other processes, such as $J/\psi$ production in $p+p$ \cite{Bodwin:2014gia, Bodwin:2015iua} and in jets \cite{Bain:2016rrv,Bain:2017wvk}, and so it is not surprising that we observe the same behavior. When considering the other LDME sets, the B$\&$K and Brambilla et al. predict significantly larger values for $\Otpz$ LDME, as seen in table \ref{tab: LDMEs}. Hence, the $P$-wave channel is not completely canceled by the $^3S_1^{[8]}$ mechanism when using these LDME values, resulting in the large endpoint divergences we observe. This indicates that our FJF calculation is more consistent with the Bodwin et al. LDME set, and further underscores the strong need for a more precise extraction of the $\psi(2S)$ LDMEs. {As far as the origin of endpoint behavior, it essentially is a feature of fixed-order calculations in this region. Such endpoint singularities might be expected to be regulated by threshold resummation, but this is out of the scope of current work.}



\section{Conclusion}
\label{sec:Conclusion}
%
In this analysis, we have used the GFIP and FJF formalisms to study the production of $\psi(2S)$ mesons within a jet. In both frameworks, \textsc{MadGraph} is used to generate distributions of hard partons from the collision with large $p_T$ corresponding to $p_T({\text{jet}})$. In the GFIP formalism, the hard parton is then showered down to the scale $2m_c$ using Pythia, at which point, the distributions are convolved with fragmentation functions for partons fragmenting into a $\psi$(2S) meson. These fragmentation functions are computed by matching onto color singlet and color octet NRQCD LDMEs. In the FJF approach, the analysis is similar except that the showering from $p_T$(jet) to $2m_c$ is computed analytically by solving the DGLAP evolution equations for the fragmentation functions. We showed that both of these approaches can successfully reproduce recent in-jet $\psi(2S)$ production data from the LHCb collaboration \cite{LHCb:2024ybz}. In our analysis, we compare with data from five $p_T^{\text{jet}}$-bins from the LHCb collaboration, over a range of jet transverse momentum values from 15 - 60 GeV/c. Our results show that these frameworks offer a significant improvement over the predictions from a naive implementation of NRQCD in \textsc{Pythia} 8 \cite{LHCb:2024ybz}. This suggests that the fragmenting jet picture, which is used by both the GFIP and FJF formalisms, is the correct way to study quarkonium production in jets.

One purpose of this work is to point out that the study of $\psi(2S)$ production in jets provides a powerful tool to discern between different extractions of the $\psi(2S)$ LDMEs. We compare predictions using LDME values from refs. \cite{Bodwin:2015iua, Butenschoen:2022qka, Brambilla:2022ayc} against the LHCb collaboration's measurements. We find that, qualitatively, the results from both the GFIP and FJF methods show the best agreement with experimental data when using the LDME set of Bodwin et al. The GFIP results based on the LDMEs from B\&K and Brambilla et al. also exhibit agreement with the data except for the highest $p_T^{\text{jet}}$-bin. However, we observe that these FJF results deviate significantly from both the data and the GFIP predictions as $z\to1$. From this analysis, we find that it is necessary to constrain the $\psi$(2S) LDMEs further, as none of the extracted LDME values provide perfect agreement with the in-jet data for all $p_T^{\text{jet}}$-bins. To find better agreement, one could even include the LHCb's in-jet production data with fits to the world's $\psi(2S)$ production data in order to constrain the $\psi(2S)$'s LDMEs further. The fragmenting jet formalisms used in this paper pave the way to make such an analysis possible. We leave this effort for future studies.\\

\section*{Note added:}
While finalizing this manuscript, Ref.~\cite{Wang:2025drz} appeared as a preprint. Their results are also obtained using the FJF formalism, incorporating threshold resummation. In contrast, our study presents calculations based on FJF without threshold resummation. {Our results within the FJF framework exhibit a significant enhancement near the endpoint region; one may expect that threshold resummation could cure this singular behavior. } We also present an additional calculation using the GFIP framework.

\section*{Acknowledgements}
We dedicate this paper to the cherished memory of Tom Mehen. Y.F. and J.R. were supported by the U.S. Department of Energy through the Topical Collaboration in Nuclear Theory on Heavy-Flavor Theory (HEFTY) for QCD Matter under award no. DE-SC0023547.  M.C. was supported by the National Science Foundation Graduate Research Fellowship under Grant No.~DGE 2139754. M.C., Y.F., and J.R. were also supported by the grant DE-FG02-05ER41367 from the U.S. Department of Energy, Office of Science, Nuclear Physics. L.D. was supported by the Guangxi Natural Science Foundation (grant No. 2026GXNSFAA00641083), and also by the Guangxi Talent Program (“Highland of Innovation Talents”).

\appendix
\section{$\psi(2S) \to J/\psi ~\pi^+ \pi^-$ decay kinematics}\label{App:A}
\subsection*{Lorentz Transformation}
Consider a Lorentz covariant four-vector, for example, a four-coordinate $(t,x,y,z)$, in a frame $F$. Now we boost this velocity to another frame $F'$. The boost velocity is $\vec{v}$, which is the velocity of frame $F'$ observed in frame $F$. The boost vector is $\vec{\beta}=\vec{v}/c = \vec{v}$(we use the natural unit c=1) and the Lorentz factor is $\gamma=\frac{1}{\sqrt{1-\vec{\beta}^2}}$ The transformation of the four-coordinate is 
\begin{equation}
\begin{split}
\left(
\begin{array}{c}
t'   \\
x' \\
y'  \\
z'  \\
\end{array}
\right)
=
\Lambda(\vec{v})
\left(
\begin{array}{c}
t   \\
x \\
y  \\
z  \\
\end{array}
\right).
\end{split}
\end{equation}
where we denote the boost matrix \begin{align}
\Lambda(\vec{v})=\Lambda(\vec{\beta})=\left( \begin{array}{cccc}
\gamma & -\gamma\beta_x & -\gamma\beta_y & -\gamma\beta_z \\
-\gamma\beta_x & 1+\frac{\gamma^2}{1+\gamma}\beta_x\beta_x &  \frac{\gamma^2}{1+\gamma}\beta_x\beta_y & \frac{\gamma^2}{1+\gamma}\beta_x\beta_z \\
-\gamma\beta_y & \frac{\gamma^2}{1+\gamma}\beta_y\beta_x &  1+\frac{\gamma^2}{1+\gamma}\beta_y\beta_y & \frac{\gamma^2}{1+\gamma}\beta_y\beta_z \\
-\gamma\beta_z & \frac{\gamma^2}{1+\gamma}\beta_z\beta_x & \frac{\gamma^2}{1+\gamma}\beta_z\beta_y &  1+\frac{\gamma^2}{1+\gamma}\beta_z\beta_z \\
\end{array} \right) 
\end{align}
and $\vec{n}=(n_x,n_y,n_z)=\frac{1}{v}(v_x,v_y,v_z)$ is the unit vector along the direction of $\vec{v}$.\\
In studying charmonium production at LHCb, we consider the $J/\psi\to \mu^+\mu^-$ in $J/\psi$'s rest frame, then boost it back to the LHCb lab frame. 

\subsection*{Two-Body Decay:$A\to B+C$}
Consider a particle $A$ with mass $M$ decaying into two particles $B$ and $C$ with masses $m_1$ and $m_2$. In the rest frame of $A$, the energy-momentum conservation reads:
\begin{align}
    M = E_1 + E_2, \ \vec{0} = \vec{p}_1 + \vec{p}_2 
\end{align}
The energies of particles $B$ and $C$ are:
\begin{align}
    E_1 = \sqrt{p^{*2} + m_1^2},\ 
    E_2 = \sqrt{p^{*2} + m_2^2}
\end{align}
where $p^{*}\equiv |\vec{p}_1| =|\vec{p}_2|$.
We can get
\begin{align}
    p^{*2} &= \frac{(M^2 - (m_1 + m_2)^2)(M^2 - (m_1 - m_2)^2)}{4M^2}
\end{align}

\subsection*{Three-Body Decay: $A\to B+C+D$}

Consider a particle $A$ with mass $M$ decaying into three particles $B$, $C$ and $D$ with masses $m_1$, $m_2$ and $m_3$. In the rest frame of $A$, the energy-momentum conservation reads:
\begin{align}
    M = E_B + E_C + E_D,\ 
    \vec{0} = \vec{p}_B + \vec{p}_C + \vec{p}_D
\end{align}
Each particle's energy is given by:

\begin{align}
    E_i &= \sqrt{p_i^2 + m_i^2}, \quad (i = B,C,D).
\end{align}
The Dalitz variables for the decaying system can be defined as:
\begin{align}
    s_{12} &= (p_B + p_C)^2, \
    s_{13} = (p_B + p_D)^2.
\end{align}
and
\begin{align}
    s_{23} &= (p_C + p_D)^2 =  M^2 + m_1^2 + m_2^2 + m_3^2 - s_{12} - s_{13}.
\end{align}
The kinematically allowed regions are
\begin{align}
    (m_i + m_j)^2 &\leq s_{ij} \leq (M - m_k)^2,
\end{align}
where $i < j$, and $i$, $j$, $k$ are mutually distinct elements of $\{1, 2, 3\}$.

We start by considering $A$ decaying into $B$ and an intermediate system $X$, which decays into $C$ and $D$: $A \to B + X,\  X \to C + D$.
The invariant mass of $X$ is $M_X^2 = (p_C + p_D)^2$. Using two-body decay formulas, we have:
\begin{align}
    p_B^* &= \frac{\sqrt{[M^2 - (m_1 + M_X)^2][M^2 - (m_1 - M_X)^2]}}{2M},
\end{align}
where $p_B^*$ is $B$'s momentum in the rest frame of $A$. Similarly, for $C$ and $D$ in the  rest frame of $X$:
\begin{align}
    p_C^* & = p_D^*= \frac{\sqrt{[M_X^2 - (m_2 + m_3)^2][M_X^2 - (m_2 - m_3)^2]}}{2M_X}.
\end{align}
To generate the magnitude of momenta of particles, one samples uniformly $M_X$ from the allowed range $m_2 + m_3 \leq M_X \leq M - m_1$.
To determine the orientation of momenta, we sample $\theta_B^*$, $\phi_B^*$ isotropically for $B$ in the rest frame of A,
and sample $\theta_C^*$, $\phi_C^*$ isotropically for $C$ and $D$ in the rest frame of $X$, yielding
\begin{align}
    p_B &= \left( \sqrt{p_B^{*2} + m_1^2}, \; p_B^* \sin\theta_B^* \cos\phi_B^*, \; p_B^* \sin\theta_B^* \sin\phi_B^*, \; p_B^* \cos\theta_B^* \right), \\
    p_C &= \left( \sqrt{p_C^{*2} + m_2^2}, \; p_C^* \sin\theta_C^* \cos\phi_C^*, \; p_C^* \sin\theta_C^* \sin\phi_C^*, \; p_C^* \cos\theta_C^* \right), \\
    p_D &= \left( \sqrt{p_C^{*2} + m_3^2}, -p_C^* \sin\theta_C^* \cos\phi_C^*, -p_C^* \sin\theta_C^* \sin\phi_C^*, -p_C^* \cos\theta_C^* \right),
\end{align}
We next need to get the momenta for particles B, C, and D in the rest frame of A. We notice that the particle B has already been sampled in the rest frame of A, so
\begin{align}
   p_B^{\text{A-rest}}= p_B = \left( \sqrt{p_B^{*2} + m_1^2}, \; p_B^* \sin\theta_B^* \cos\phi_B^*, \; p_B^* \sin\theta_B^* \sin\phi_B^*, \; p_B^* \cos\theta_B^* \right).
\end{align}
Particles C and D are sampled in the rest frame of X, whose relative momentum observed in the rest frame of A is $-\vec{p}_B^*$. Therefore, the relative velocity of X's rest frame observed in the rest frame of A is 
\begin{align}
\vec{v}_X^*=-\frac{\vec{p}_B^*}{E_X^*}= -\frac{\vec{p}_B^*}{E_C^*+E_D^*}   
\end{align} 
We boost their momenta back to the rest frame of $A$,
\begin{align}
    p_C^{\text{A-rest}} &= \Lambda(-\vec{v}_X^*) \left( \sqrt{p_C^{*2} + m_2^2}, \; p_C^* \sin\theta_C^* \cos\phi_C^*, \; p_C^* \sin\theta_C^* \sin\phi_C^*, \; p_C^* \cos\theta_C^* \right), \\
    p_D^{\text{A-rest}} &= \Lambda(-\vec{v}_X^*) \left( \sqrt{p_C^{*2} + m_3^2}, -p_C^* \sin\theta_C^* \cos\phi_C^*, -p_C^* \sin\theta_C^* \sin\phi_C^*, -p_C^* \cos\theta_C^* \right),
\end{align}
where $\Lambda(-\vec{v}_X^*)$ is the Lorentz boost from the rest frame of $X$ to the rest frame of $A$.

Finally, we need to boost the momenta of particles B, C, and D back to the lab frame. To this end, we notice that the relative momentum of the rest frame of A observed in the lab frame is $\vec{p}_A^{\text{lab}}$, which is known. Therefore, the relative momentum of the rest frame of A observed in the lab frame is 
\begin{align}
\vec{v}_A^{\text{lab}}=   \frac{\vec{p}_A^{\text{lab}}}{E_A^{\text{lab}}}
\end{align} 
We finally get the momenta of particles B, C, and D in the lab frame via
\begin{align}
p_{B,C,D}^{\text{lab}} &= \Lambda(-\vec{v}_A^{\text{lab}}) p_{B,C,D}^{\text{A-rest}} 
\end{align}
\subsection*{Four-body decay: $A\to B+C+D \to B_1+B_2 +C +D$}
Here, we further require that $ B\to B_1 + B_2$, where the mass of $B_1$ and $B_2$ are $m_{B_1}$  and $m_{B_2}$. 
Particle B has already been sampled in the rest frame of A, and then we want to deal with $ B \to B_1 + B_2$ in the rest frame of B. Using two-body decay formulas, we have
\begin{align}
    p_{B_1}^* &= \frac{\sqrt{[m_1^2 - (m_{B_1} + m_{B_2})^2][m_1^2 - (m_{B_1} - m_{B_2})^2]}}{2m_1^2},
\end{align}
where $p_{B_1}^*$ is $B_1$'s momentum in the rest frame of B. We then sample $\theta_{B_1}^*$, $\phi_{B_1}^*$ isotropically for $B_1$ in the rest frame of B, and get
\begin{align}
   p_{B_1}^{\text{B-rest}}& = \left( \sqrt{p_{B_1}^{*2} + m_1^2}, \; p_{B_1}^* \sin\theta_{B_1}^* \cos\phi_{B_1}^*, \; p_{B_1}^* \sin\theta_{B_1}^* \sin\phi_{B_1}^*, \; p_{B_1}^* \cos\theta_{B_1}^* \right)\\
p_{B_2}^{\text{B-rest}}& = \left( \sqrt{p_{B_1}^{*2} + m_2^2}, \; -p_{B_1}^* \sin\theta_{B_1}^* \cos\phi_{B_1}^*, \; -p_{B_1}^* \sin\theta_{B_1}^* \sin\phi_{B_1}^*, \; p_{B_1}^* \cos\theta_{B_1}^* \right)
\end{align}
Now, the relative momentum of $B$ observed in the rest frame of A is $p_B^{\text{A-rest}}$, meaning that the relative momentum of $B$ observed in the rest frame of A is 
\begin{align}
\vec{v}_B^{\text{A-rest}}=\frac{\vec{p}_B^{\text{A-rest}}}{E_B^{\text{A-rest}}} = \vec{v}_B^* = \frac{\vec{p}_B^*}{E_B^*}
\end{align}
We can therefore firstly boost the B's momentum to the rest frame of A by
\begin{align}
   p_{B_1}^{\text{A-rest}}=\Lambda(-\vec{v}_B^{\text{A-rest}}) p_{B_1}^{\text{B-rest}},\ 
   p_{B_2}^{\text{A-rest}}=\Lambda(-\vec{v}_B^{\text{A-rest}}) p_{B_2}^{\text{B-rest}}
\end{align}
Then, since the relative momentum of A observed in the lab frame is $\vec{p}_A^{\text{lab}}$, which is known, we can finally boost the momenta of $B_{1,2}$ back to the lab frame by
\begin{align}
p_{B_1}^{\text{lab}}=\Lambda(-\vec{v}_A^{\text{lab}}) p_{B_1}^{\text{A-rest}},\  p_{B_2}^{\text{lab}}=\Lambda(-\vec{v}_A^{\text{lab}}) p_{B_2}^{\text{A-rest}} 
\end{align}

\subsection*{Application to $\psi(2S)\to J/\psi + \pi^+ + \pi^- \to \mu^+ + \mu^- + \pi^+ + \pi^- $}
We identify $A \implies \psi(2S), B_1 \implies \mu^+, B_2 \implies \mu^-, C \implies \pi^+, D \implies \pi^-$.
We need to sample the following kinematics quantities uniformly,
\begin{align}
    M_X & \in [ m_{2} +m_{3}, M - m_1 ] \\
    \cos\theta_C^*, \cos\theta_{B_1}^* &\in [-1,1] \\
    \phi_C^*, \phi_{B_1}^* &\in [0,2\pi), 
\end{align}
and calculate $\vec{v}_X^*$, $\vec{v}_B^{\text{A-rest}}$ and $\vec{v}_A^{\text{lab}}$;  Applying all kinds of boosting, we can get $p_{B_1}^{\text{lab}}$, $p_{B_2}^{\text{lab}}$, $p_{C}^{\text{lab}}$, $p_{D}^{\text{lab}}$, i.e., the measured momenta of $\mu^+$, $\mu^-$, $\pi^+$ and $\pi^-$ in lab frame.

\bibliographystyle{JHEP}
\bibliography{biblio}

\end{document}